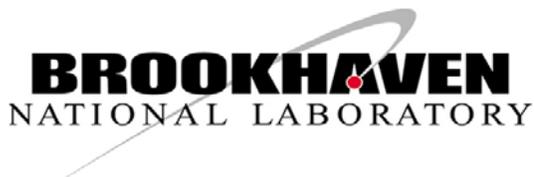

BNL-94868-2011-CP

# NSLS-II X-RAY DIAGNOSTICS DEVELOPMENT

*P. Ilinski*

## DISCLAIMER

# NSLS-II X-RAY DIAGNOSTICS DEVELOPMENT*

P. Ilinski#, BNL, Upton, NY 11973, U.S.A.

## Abstract

NSLS-II x-ray diagnostics will provide continuous on-line data of electron beam dimensions, which will be used to derive electron beam emittance and energy spread. It will also provide information of electron beam tilt for coupling evaluation.

X-ray diagnostics will be based on imaging of bending magnet and three-pole wiggler synchrotron radiation sources. Diagnostics from three-pole wiggler source will be used to derive particles energy spread. Beta and dispersion functions will have to be evaluated for emittance and particles energy spread calculations. Due to small vertical source sizes imaging need to be performed in x-ray energy range. X-ray optics with high numerical aperture, such as compound refractive lens, will be used to achieve required spatial resolution. Optical setups with different magnifications in horizontal and vertical directions fill be employed to deal with large aspect ratio of the source. X-ray diagnostics setup will include x-ray imaging optics, monochromatization, x-ray imaging and recording components.

## PERFORMANCE REQUIREMENTS

X-ray diagnostics should provide imaging of the radiation sources in the range of 5 mA to 500 mA of stored electron beam current. Expected electron beam parameters are: horizontal emittance from 0.6 to 2 nm, vertical emittance ~8 pm, and the energy spread from 0.05% to 0.1%. The electron beam vertical size at the diagnostics sources locations is 12-14 μm, Table 1. This will require spatial resolution of the diagnostics to be better then three microns in order to observe changes of the vertical beam size. Field of view of the imaging system need be two to four times of the imaged source.

Table 1: Radiation sources for x-ray diagnostics.

|  | BM | 3-pole wiggler |
|---|---|---|
| $\sigma_x$ [μm] | 75 | 136 |
| $\sigma_y$ [μm] | 14 | 12 |
| $\sigma_{x'}$ [μrad] | 56 | 77 |
| $\sigma_{y'}$ [μrad] | 0.8 | 0.9 |
| $B$ [T] | 0.4 | 1.14 |
| $E_c$ [keV] | 2.4 | 6.8 |
| $\sigma'_{ph} \sim 1/\gamma$ [μrad] | 170 | 170 |
| $\sigma_{ph}$ [μm] | 0.24 | 0.09 |

## RADIATION SOURCES

The x-ray diagnostics will utilize BM-A bending magnet and three-pole wiggler located at Sector 22. Radiation source from BM-A bending magnet is located at the beginning of the dipole where the dispersion function is negligible. The dispersion function has substantial value at the three-pole-wiggler location, which will contribute to the source size. Particles energy spread can be derived from measurements of source sizes at locations with different values of the dispersion function. Electron beam sizes without and with three and eight damping wigglers are shown in Fig. 1.

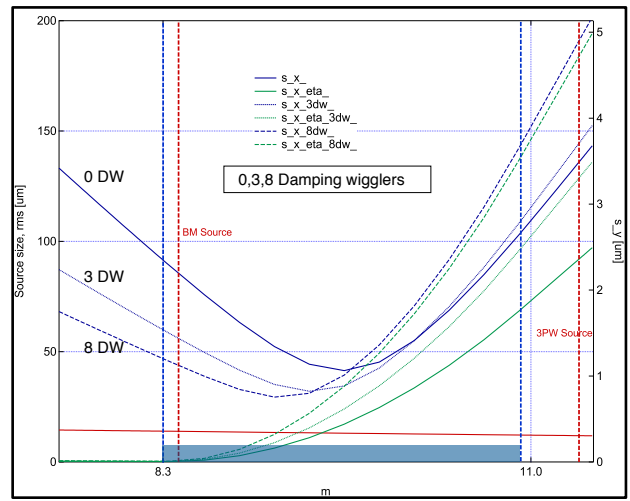

Fig. 1: Electron beam horizontal source sizes at bending magnet and three-pole-wiggler locations. Total rms source size (blue), contribution due to the dispersion (green). Damping wigglers: without (solid), 3 (dots), 8 (dash).

## SPATIAL RESOLUTION OF THE OPTICAL SYSTEM

Spatial resolution of the system is a convolution of resolutions of the optical components: x-ray optics, transport system and detection. Resolution of x-ray optics is defined by diffraction and aberrations; beam transport may introduce scattering from media and aberrations from mirror or crystal imperfections; detection resolution combines resolutions of scintillator screen, micro-objective and CCD detection camera. Dependence of the optical system components spatial resolution on setup parameters is listed in Table. 2. Optimization of the diagnostics setup requires matching resolution of all optical components.

Table 2: Spatial resolution of the optical components.

| Optical Element | X-ray optics | Scintillator | Objective | CCD |
|---|---|---|---|---|
| Resolution | $\lambda_{x\text{-}ray}/2\cdot NA$ | 1/thickness | $\lambda_{visible}/2\cdot NA_{obj}$ | pixel / Magnif. |

___________________________________________
*Work supported by DOE contract DE-AC02-98CH10886
#pilinski@bnl.gov

# RESOLUTION OF X-RAY OPTICS

Compound refractive lens (CRL) x-ray optics will be used to achieve required high spatial resolution, which is possible due to high CRL numerical aperture (NA). A pinhole optics setup will be employed along with the CRL setup due to its robustness. Both setups were optimized to achieve required performance for given conditions.

## Pinhole optics

Pinhole optics does not require monochromatization of the photon beam, photon energy filtering will be adequate to achieve moderate resolution. Along with its simplicity, pinhole optics has limited spatial resolution due to blurring effect. Spatial resolution does not increase with increase of pinhole numerical aperture, best resolution of the pinhole setup corresponds to the Fresnel number of $F=0.7$ [1-2].

Spatial distribution of two-dimensional pinhole Point Spread Function (PSF) can be quite complicated, which prevents to perform deconvolution of the resulting image reliably. PSF profiles of square pinholes from 5 to 100 μm calculated with SRW [3] are shown at Fig. 2. Spatial resolution of ~10-15 microns can be feasible to achieve with pinhole setup in the x-ray energy range.

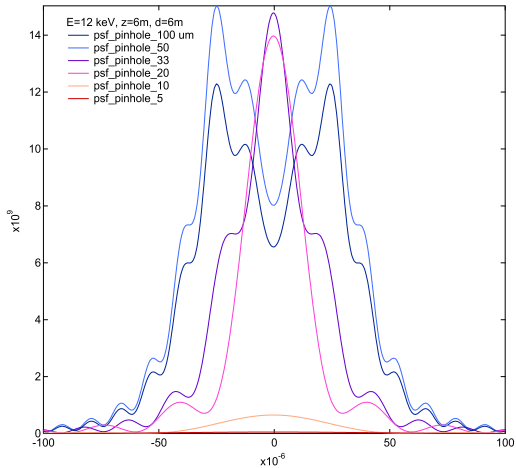

Fig. 2: PSF profiles of 10, 20, 33 ($F=0.7$) 50 and 100 μm square pinholes.

## Optimization of the pinhole setup

Pinhole distance, $z$, from the source and pinhole size, $2a$, can be defined for an optimal pinhole setup Fresnel number, $F=0.7$, and for required PSF, Fig. 3. For best resolution pinhole should be located close to the source, nevertheless closest location of the first optical element depends on the vacuum chamber design and girder layout. Moreover, x-ray pinholes or slits with dimensions less then 20 μm are not practical, which also prevents to achieve high spatial resolution with pinhole setup.

Spatial resolution of the pinhole setup will also improve with increase of the setup total length, but dependence is very gradual, Fig. 3. In our case, the required spatial resolution can be achieved for pinhole setup, which will be located within the storage ring tunnel.

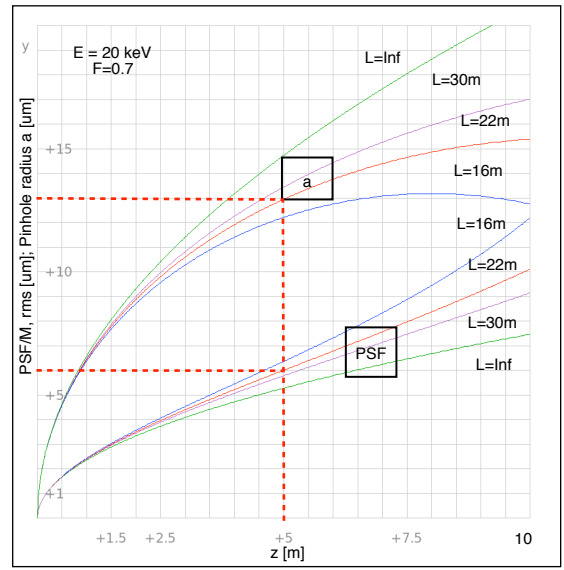

Fig.3: Optimization of the pinhole setup. Pinhole PSF and pinhole half-size, $a$, are shown for different length of the pinhole setup versus distance from the source to the pinhole, $z$.

## Compound Refractive Lens optics

An x-ray imaging setup with CRL optics will be employed in order to achieve spatial resolution of 1-2 microns. CRL x-ray optics was successfully implemented at PETRA-III diagnostics beamline for emittance measurements [4].

Compound refractive lens is a set of x-ray refractive lenses, number of single lenses are stack together to achieve required total focal length [5]. CRL optics has inline geometry, which is easy to align, and it is commercially available. Proper monochromatization is required because of CRL high chromatic aberrations.

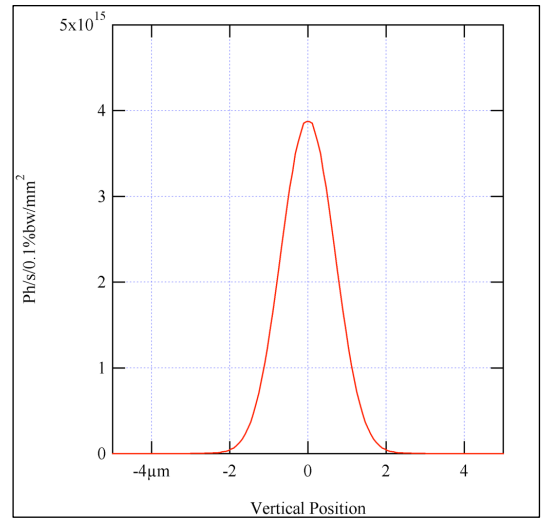

Fig. 4: CRL vertical PSF.

Vertical and horizontal CRL PSFs calculated with SRW [3] are presented at Fig. 4 and 5 correspondently. Resolution in vertical direction is limited due to narrow vertical beam divergence, which in this case defines the CRL NA. The NA in horizontal plane is defined by the total x-ray absorption of CRL material, which is beryllium. Horizontal PSF is non-symmetrical due to non-symmetry of the bending magnet synchrotron radiation source.

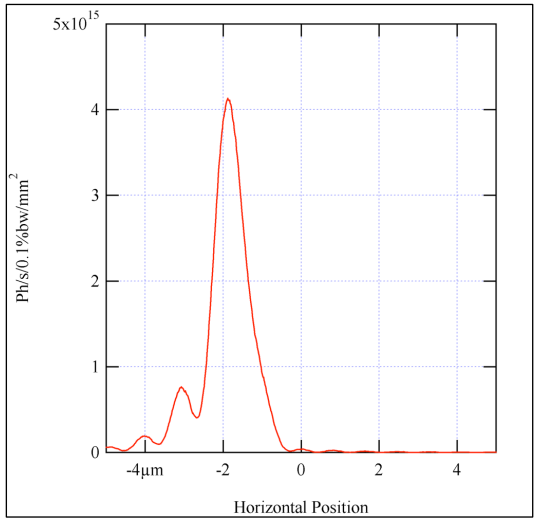

Fig. 5: CRL horizontal PSF.

*Resolution of detection system*

Detection system consists of scintillation screen, micro-objective and CCD-camera. Resolution of components should match each other to minimize the total resolution of the detection system. To achieve resolution of few microns a thin, ~10-µm-thick scintillation crystal and 10x micro-objective will be used.

## BEAMLINE LAYOUT

The layout of proposed diagnostics beamline is shown at Fig. 6. Each of two diagnostics beamlines will employ two kinds of x-ray optics setups: pinhole and CRL. A multilayer mirror or a single crystal monochromator will be used to monochromatize incident radiation. Pinhole and CRL setups will be interchangeable.

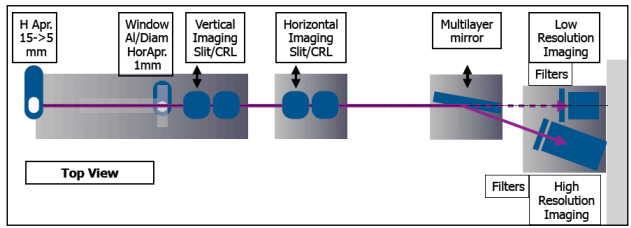

Fig. 6: General layout of x-ray diagnostics beamline.

Table 3: Aspect ratio of radiation sources.

|  | BM | BM 3DW | BM 8DW | 3PW | 3PW 3DW | 3PW 8DW |
|---|---|---|---|---|---|---|
| Source size V [µm] | 14 | 14 | 14 | 12 | 12 | 12 |
| Source size H [µm] | 75 | 49 | 39 | 133 | 137 | 138 |
| Aspect ratio H/V | 5 | 4 | 3 | 11 | 11 | 12 |

Horizontal and vertical source dimensions are quite differ, the resulting image will have high aspect ratio in case when magnification in horizontal and vertical directions are the same, Table 3. This will require CCD-camera with large number of pixels to record image with proper sampling. In order to fit image in horizontal plane and to achieve required resolution of the detection system in vertical plane, horizontal and vertical imaging will be decoupled. Slit and one-dimensional CRL will be used for imaging in each direction.

## SPECTRAL FLUX

Spectral flux from bending magnet is sown at Fig. 7 before and after attenuation in 1-mm-thick aluminum window and 10-m-long air path. X-ray diagnostics will be setup in air, which will simplify the design and maintenance. For optimal resolution and flux performance operation energy will be set at 20 keV. X-ray radiation from the source is imaged onto the fluorescent crystal, which converts x-rays into visible light photons, visible light imaged then onto the CCD chip by micro-objective. Calculations of total flux budget at the image plane of the CCD-camera shows that x-ray diagnostics will be able to operate at 5 mA. Attenuation filters will be used to accommodate dynamic range from 5 to 500 mA of stored electron beam current.

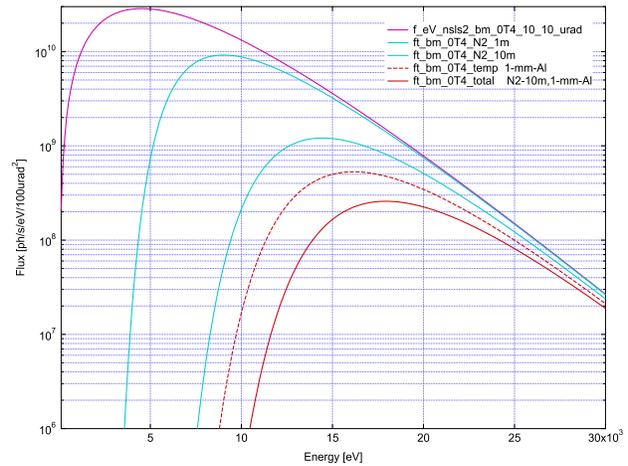

Fig. 7: Bending magnet Spectral flux, 500 mA, 10x10µrad$^2$ (purple); attenuated by 1-mm-thick aluminum window and 10 m air path (solid red).